\documentclass[12pt, hyper]{article}
\usepackage{amsmath,amsopn,amssymb,amsfonts}
\usepackage{amsthm}
\usepackage{hyperref}
\usepackage[pdftex]{graphicx}
\usepackage{epsfig}
\usepackage{mathrsfs}
\DeclareMathAlphabet{\mathpzc}{OT1}{pzc}{m}{it}
\usepackage{verbatim}
\usepackage{multirow}
\usepackage{tikz}
\usepackage{pgfplots}
\usepackage{arydshln}
\usepackage{cite}
\usepackage{amsthm}
\usepackage[margin=1cm]{caption}
\usepackage[font={small}]{caption}

\def\CA{{\cal A}}\def\CB{{\cal B}}
\def\CF{{\cal F}}\def\CG{{\cal G}}
\def\CI{{\cal I}}

\def\CO{{\cal O}}\def\CP{{\cal P}}
\def\CR{{\cal R}}

\def\CC{{\cal C}}

\def\a{\alpha}
\def\d{\delta}

\def\t{\tau}
\def\w{\omega}

\usepackage{cleveref}

\begin{document}

\begin{titlepage}
\vfill
\begin{flushright}
{\tt\normalsize KIAS-P16077}\\
\end{flushright}
\vfill
\begin{center}
{\Large\bf Bootstrapping Pure Quantum Gravity in AdS$_3$
}

\vfill

Jin-Beom Bae, Kimyeong Lee, and Sungjay Lee

\vskip 5mm
{\it Korea Institute for Advanced Study \\
85 Hoegiro, Dongdaemun-Gu, Seoul 02455, Korea}

\end{center}
\vfill

\begin{abstract}
\noindent
The three-dimensional pure quantum gravity with negative cosmological constant is supposed to be dual to
the extremal conformal field theory of central charge $c=24k$ in two dimensions.
We employ the conformal bootstrap method to analyze the extremal CFTs, and find numerical evidence
for the non-existence of the extremal CFTs for sufficiently large central charge ($k \ge 20$).
We also explore near-extremal CFTs, a small modification of extremal ones, and
find similar evidence for their non-existence for large central charge.
This indicates, under the assumption of holomorphic factorization,
the pure gravity in the weakly curved AdS$_3$ do not exist as a consistent quantum theory.
\end{abstract}

\vfill
\end{titlepage}

\parskip 0.1 cm
\renewcommand{\thefootnote}{\#\arabic{footnote}}
\setcounter{footnote}{0}

\parskip 0.2 cm

\section{Introduction}

AdS/CFT correspondence has provided many new insights to the nature of quantum gravity.
As a simple example of the correspondence, the three-dimensional pure gravity with negative cosmological constant
\begin{align} \label{Igrav}
  \CI_\text{grav} = \frac{1}{16\pi G_N} \int d^3x \sqrt{g} ~ \Big[ \CR + \frac{2}{\ell^2}  \Big]\ ,
\end{align}
where $\ell$ denotes the AdS length scale, has received constant attention for many years.
It was proposed in \cite{Witten:2007kt}
that the partition function of the pure quantum gravity in AdS$_3$
should be holomorphically factorized.
Under this assumption, the  two-dimensional dual conformal field theory (CFT) can be described as a product of a holomorphic CFT
and an anti-holomorphic CFT of central
charges  $c_L=c_R=24k$ with positive integer $k$.
In the present work, we  focus mostly on the holomorphic CFT.

The pure gravity theory in AdS$_3$ has two kinds of excitations.
One of them is made of the boundary gravitons and the other one of the BTZ black holes.
In the dual (holomorphic) CFT, the former can be identified as
Virasoro descendants of the vacuum
while the latter as the Virasoro primaries.
Among the spectrum, the lowest primary states are
corresponding to the lightest BTZ black hole in AdS$_3$ of finite-size horizon. The modular invariance and the Brown-Heanneaux formula then implies that conformal weight of the lowest primary is given by $k+1$\cite{Witten:2007kt}.

The partition function $Z_k(\t)$
of the dual holomorphic CFT can be expanded as
\begin{align} \label{ECFT_partition}
  Z_k(\t) = q^{-k} \left[ \prod_{n=2}^\infty \frac{1}{1-q^n} + \CO(q^{k+1})\right] \
\end{align}
around small $q=e^{2\pi i\t}$.
The first term in (\ref{ECFT_partition}) counts the contribution of Virasoro vacuum descendants,
and is not modular invariant by itself.
This naturally requires the additional contribution of Virasoro primaries, captured by $\CO(q^{k+1})$.
Together they define the holomorphic part of the pure quantum gravity partition function in AdS$_3$.
Such (holomorphic) CFTs consistent with the above conditions have been called
the extremal CFTs.

The extremal CFT for $k=1$ is the famous monster CFT, constructed
by Frenkel, Lepowski and Meurman\cite{Frenkel:1988xz} as a $\mathbb{Z}_2$ orbifold
of free bosons on the Leech lattice. The symmetry of this CFT is the monster group, that is, the largest sporadic finite simple group.
However, it is not entirely clear whether there exist any consistent extremal CFTs with $k \geq 2$. Despite
an early attempt to prove their non-existence for
sufficiently large $k$ ($k\geq42$) using the modular
differential equation\cite{Gaberdiel:2007ve}, it was shown in \cite{Gaiotto:2008jt} that there is a loophole in that argument, and the investigation of this issue has remained inconclusive.

In this present work, we explore the existence of the extremal CFTs
for $k\geq 2$ using conformal bootstrap method. The conformal bootstrap
method is a program to utilize general consistency
conditions such as unitarity and associativity of the operator
product expansions (OPE) to understand the nonperturbative
aspects of CFTs. The program usually considers four-point
correlation functions of bosonic operators. Given certain simple
assumptions about the spectrum, the crossing symmetry of four-point
functions provides nontrivial constraints, called the bootstrap equations.

In recent years the numerical approach to the conformal bootstrap\cite{Rattazzi:2008pe,Poland:2011ey} has led to remarkable progress in understanding of the CFTs
in higher dimensions, for example, critical exponents, central charges, and
OPE coefficients of three-dimensional Ising model\cite{ElShowk:2012ht, El-Showk:2014dwa, Kos:2014bka} and $O(N)$ vector models\cite{Kos:2013tga, Kos:2015mba}.
In 2D CFT, the recursive representation of the Virasoro conformal block\cite{Zamolodchikov:1985ie} has been implemented in
the numerical conformal bootstrap. This led to a number of new constraints on both spectrum\cite{Lin:2015wcg} and OPE coefficients\cite{Esterlis:2016psv}, previously inaccessible by analytic methods.

In our numerical approach to the extremal CFTs for $k\geq 2$  we focus on the four-point correlation function of a chiral primary operators $\phi(z)$ of the lowest conformal weight $h_{\phi} = k+1$.
We found numerical evidence supporting  the non-existence of extremal CFTs for $k \ge 20$.
This implies that, under the assumption of holomorphic factorization, the pure quantum gravity
may not exist in the weakly curved AdS$_3$ space.

We also analyzed the near-extremal CFT\cite{Benjamin:2016aww}, a modification of extremal CFT to include small black holes of zero classical entropy. Our analysis indicates that the near-extremal CFT would not be consistently realized for $k \ge 4$.

The rest of the paper is organized as follows. Section 2 includes brief reviews on extremal conformal field theory and the Zamolodchikov recursive relation of Virasoro conformal block.
Section 3 summarizes the numerical bootstrap process. In section 4, we provides our settings and numerical results. Finally, section 5 conclude with summary of numerical results and discussions.


\section{Brief Reviews}

\subsection{Extremal Conformal Field Theory}

Three-dimensional pure gravity with negative cosmological constant is described by the Einstein-Hilbert action (\ref{Igrav}).
One can also describe the pure gravity in three dimensions as
$SL(2,\mathbb{R})\times SL(2,\mathbb{R})$ Chern-Simons gauge theory at least classically,
\begin{align} \label{CSaction}
  \CI_\text{grav} = \frac{k_L}{4\pi} \CI_\text{CS}\left(\CA_L\right) - \frac{k_R}{4\pi} \CI_\text{CS}\left(\CA_R \right)
\end{align}
with
\begin{align}
  \CI_\text{CS}\left(\CA\right) = \int
  \mbox{Tr} \left[ \CA \wedge d \CA  + \frac{2}{3}   \CA \wedge \CA \wedge \CA \right]\ .
  \nonumber
\end{align}
Here the $SL(2,\mathbb{R})$ gauge fields $\mathcal{A}_L$ and $\mathcal{A}_R$ are defined by
\begin{equation}
  \mathcal{A}_L = \omega - \frac{e}{\ell}, \qquad \mathcal{A}_R = \omega + \frac{e}{\ell}\ ,
\end{equation}
where $e$ and $\w$ denote the vierbein and the spin connection. The two Chern-Simons
levels $k_L$ and $k_R$ have to be quantized to integer values in order to make the action (\ref{CSaction}) gauge invariant.
One can show that the action (\ref{CSaction}) agrees with the Einstein-Hilbert action (\ref{Igrav})
without the gravitational Chern-Simons coupling, when we set
$k_L = k_R =k= {\ell}/{16G_N}$.

Combined with the Brown-Henneaux formula $c=\frac{3\ell}{2G_N}$\cite{Brown:1986nw}, the quantization of the Chern-Simons level suggests that
the dual conformal field theory defined on the boundary of AdS$_3$ has the central charge
\begin{align}
  c_L = c_R = 24 k
\end{align}
where $k$ is an arbitrary positive integer.
Guided by the above observation, it has been proposed in \cite{Witten:2007kt} that
the three-dimensional pure quantum gravity in AdS$_3$ can be defined as
a two-dimensional unitary conformal field theory with central charge $c_L=c_R=24k$.
For these values of the central charge,
a dual conformal field theory can factorize into a holomorphic CFT and an anti-holomorphic CFT.
Due to the modular invariance,
the holomorphic (anti-holomorphic) CFT  with $c_L=24k$ ($c_R=24k$) is further restricted to have
operators of integer-valued conformal weights.
It implies that the pure quantum gravity in AdS$_3$
has no fermionic states and the parity symmetry remains
intact at the quantum level, so does the dual CFT.

In the present work, we will focus on the holomorphic part of the dual parity-preserving CFT, called extremal CFT.
As discussed in the introduction, the partition function
of the extremal CFT is as close as possible to the
vacuum Virasoro character corresponding to the boundary gravitons.
More precisely, the modular invariant partition function
has to be of the form
\begin{align}
  Z(\t) = \mbox{Tr}\left[ q^{L_0 -\frac{c}{24}} \right]=
  q^{-k} \left[ \prod_{n=2}^\infty \frac{1}{1-q^n} + \CO(q^{k+1}) \right]\ ,
\end{align}
which includes the contribution from the primaries above the vacuum.
The primary operators of conformal weight larger than $k$ ($h\geq k+1$) are interpreted
as states corresponding to the BTZ black holes with (classical) finite-size horizon.
It is a key property that the extremal CFT have a sparse spectrum of states at low energies
and a large gap from the vacuum to the first primary operator.

Due to the modular invariance, one can also express the above partition
function $Z(\t)$ having a pole of order $k$ only at $q=0$ as a polynomial in
the $J(\t)$ function of order $k$. For example, the partition function of the $k=1$ extremal CFT is,
\begin{equation}
Z(\t)=J(\t) = \frac{1}{q}+196884 q+21493760 q^2+864299970 q^3+20245856256 q^4+ \mathcal{O}(q^5) .
\end{equation}

For the case of $k=1$, the extremal CFT can be constructed as a $\mathbb{Z}_2$ orbifold of free bosons on the Leech lattice \cite{Frenkel:1988xz}.
Among $71$ holomorphic CFTs with $c=24$ found in \cite{Schellekens:1992db}, this is the only one without any Kac-Moody algebra. That is expected for a CFT dual to the pure gravity in AdS$_3$. However, the investigation to the existence of the extremal CFT with $k \geq 2$ has remained incomplete.


\subsection{Virasoro Conformal Block}

Let us consider a four-point correlation function
of a chiral primary operator $\phi(x)$ of conformal weight $(h,\bar h)=(h_\phi,0)$
in a holomorphic part of a unitary parity-preserving CFT with $c_L=c_R=24k$,
\begin{align}\label{4point}
  \Big\langle \phi(z_1) \phi(z_2) \phi(z_3) \phi(z_4) \Big\rangle\ .
\end{align}
Using the global conformal symmetry $SL(2,\mathbb{R})$, we can insert the operator $\phi(z)$ at
$z_1 = 0, z_2 = z, z_3 = 1$, and $z_4 = \infty$. The complex parameter $z$ governing the dynamics
of the correlation function is often called the conformal cross ratio.

The operator product expansion (OPE) of the chiral operator $\phi(z)$ is
\begin{align}
  \phi(z) \phi(0) = \sum_\CO C_{\phi\phi \CO} z^{h_\CO -2h_\phi} \hat \CO(z)\ ,
\end{align}
where the sum is over Virasoro primary operators $\CO$ of conformal weight $h_\CO$, and
$\hat \CO(z)$ includes the primary operator $\CO$ and its Virasoro descendants. Note that
the OPE coefficients $C_{\phi\phi\CO}$ are real-valued for any unitary parity-preserving CFTs \cite{Rattazzi:2008pe}.
Using the OPE between pairs of the operator $\phi$, one can expand the four-point correlation function
(\ref{4point}) of a holomorphic CFT into the following form
\begin{align}\label{fourpointexpansion}
 \Big\langle \phi(0) \phi(z) \phi(1) \phi(\infty) \Big\rangle =
 \sum_\CO C_{\phi\phi\CO}^2 \CF\left(c,h_\phi,h_\CO;z\right)\ .
\end{align}
The function $\CF(c,h_\phi,h_\CO;z)$ is the Virasoro conformal block. It is known that
the global conformal block in two dimensions can be expressed in terms of hypergeometric
functions. However the closed form of the Virasoro conformal block still remains unknown due
to the fact that the local Virasoro generators $L_n$ spans a much larger space of states than
that of global conformal generators.

Nevertheless the two-dimensional Virasoro algebra suggests a remarkable
recursive relation for the Virasoro conformal blocks, proposed by Zamolodchikov\cite{Zamolodchikov:1985ie,Perlmutter:2015iya}.
To elaborate it, we start with the elliptic representation of the Virasoro block
\begin{align}
  \CF\big(c,h_\phi ,h_\CO; z) =
  \big[z(1-z)\big]^{\frac{c-1}{24}-2h_\phi}
  \big[ 16 \ \mathpzc{q}(z)\big]^{h_\CO-\frac{c-1}{24}} \vartheta_3\big(\mathpzc{q}(z)\big)^{\frac{c-1}{2}-16 h_\phi}
  H\big(c,h_\phi,h_\CO; z)\ ,
\end{align}
where the elliptic variable $\mathpzc{q}(z)$ is defined by
\begin{align}
  \mathpzc{q}(z) = e^{\pi i \tau(z)} \ , \qquad
  \tau(z) =  i \cdot \frac{{_2F_1}(\frac12,\frac12,1;1-z)}{{_2F_1}\big(\frac12,\frac12,1;z\big)}\ ,
\end{align}
and the Jacobi theta function is $\vartheta_3(\mathpzc{q}(z))=\sum_{n\in \mathbb{Z}} \mathpzc{q}(z)^{n^2}$.
The rational function $H(c,h_\phi,h_\CO,\mathpzc{q}(z))$ of $h_\CO$ has a recursion formula
\begin{align} \label{recursive}
  H(c,h_\phi,h_\CO, \mathpzc{q}(z)) = 1 + \sum_{m,n \geq 1} \frac{(16 \ \mathpzc{q}(z))^{mn} R_{m,n}(c,h_\phi)}{h_\CO - h_{m,n}(c)}
  H(c,h_\phi, h_{m,n}(c)+m n,\mathpzc{q}(z))\ .
\end{align}
This accounts for the fact that the Virasoro block has poles at
$h_\CO=h_{m,n}(c)$
\begin{align}\label{nullweight}
  h_{m,n}(c) = \frac{c-1}{24} + \frac{(\alpha_+ m + \alpha_- n)^2}{4}\ , \qquad
  \alpha_{\pm} =  \sqrt{\frac{1-c}{24}} \pm \sqrt{\frac{25-c}{24}}\ ,
\end{align}
due to the null states at the level $mn$. Since the level $mn$ null
state can generate its own Verma module, the residue at $h_\CO=h_{m,n}(c)$
has to be proportional to the Virasoro block with conformal weight $h_\CO=h_{m,n}(c)+mn$.

From the requirement that the residue vanishes when
the central charge $c$ is the minimal model central charge,
one can carefully determine the remaining piece \cite{Zamolodchikov:1985ie,Perlmutter:2015iya},
\begin{align}
  R_{m,n}(c,h_\phi) = -\frac{1}{2} \prod_{r,s} \left( 2 \ell_{h_\phi} - \frac{\ell_{r,s}}{2}\right)^2 \left( \frac{\ell_{r,s}}{2}\right)^2
  \times \prod_{a,b}  \left( \frac{1}{\ell_{a,b}} \right)
\end{align}
with
\begin{align}
  \ell_{h_\phi} = \sqrt{h_\phi + \frac{1-c}{24}}\ , \qquad
  \ell_{r, s} = r \alpha_+ + s \alpha_-\ .
\end{align}
The products run over
\begin{align}
r &= -m+1, -m+3, \cdots , m-3, m-1 \nonumber \\
s &= -n+1, -n+3, \cdots , n-3, n-1\nonumber \\
a &= -m+1, -m+2, \cdots , m \nonumber \\
b &= -n+1, -n+2, \cdots , n\ ,
\end{align}
but exclude the two cases $(a,b) = (0 , 0)$ and $(a,b) = (m , n)$.


\section{The Numerical Virasoro Bootstrap}

The conformal bootstrap program utilizes the basic principles of quantum theory such as
associativity of the OPE and unitarity. Using the OPE between different pairs of
operators, one can have different expansions of the same four-point correlation
function (\ref{fourpointexpansion}). Crossing symmetry of the 4-point function which is the consquence of the associativity of the OPE, then leads to a constraint equation,
\begin{align}\label{bootstrap1}
  \sum_\CO C_{\phi\phi\CO}^2 \Big\{ \CF(c,h_\phi,h_\CO;z) -
  \CF(c,h_\phi,h_\CO;1-z) \Big\} = 0\ .
\end{align}
For a compact expression of the constraint equation, let us introduce a
function
\begin{align}
  \CG(c,h_\phi,h_\CO;z) =
  \CF(c,h_\phi,h_\CO;z) -
  \CF(c,h_\phi,h_\CO;1-z)\ .
\end{align}

Recent breakthroughs in the conformal bootstrap rely on translating
the above constraint equation into a numerical problem\cite{Poland:2011ey}, instead of solving
(\ref{bootstrap1}) exactly. The basic idea is to find a linear
functional $\a$ that positively acts on the vacuum Virasoro block and
non-negatively acts on all the other Virasoro blocks,
\begin{align}\label{constraint}
  \a\Big[ \CG(c,h_\phi,0;z) \Big] > 0\ , \qquad
  \a\Big[ \CG(c,h_\phi,h_\CO;z) \Big] \geq 0\ \ (^\forall \CO \neq \bf{1}),
\end{align}
under a certain assumption on CFT data, for instance, $\{c,h_\phi,h_\CO\}$ where ${\bf 1}$ denotes the unit operator.
If one can find such a functional $\a$, then
\begin{align} \label{feasible_problem}
  C_{\phi\phi{\bf 1}}^2 \a\Big[ \CG(c,h_\phi,0;z)\Big]
  + \sum_{\CO\neq {\bf 1}} C_{\phi\phi\CO}^2\a\Big[ \CG(c,h_\phi,h_\CO;z)\Big] > 0\ ,
\end{align}
because, as discussed in section 2.2,
$C_{\phi\phi{\bf 1}}^2>0$ and $C_{\phi\phi\CO}^2\geq0$ for the CFTs of our interests\footnote{A few
nontrivial OPE coefficients of the extremal CFTs for $k=1,2$ were computed in \cite{Gaiotto:2007xh}.}.
Applying this functional $\a$ on the both sides of (\ref{bootstrap1}), we find a contradiction. This
implies that there is no consistent CFT with the given data, say $\{c,h_\phi,h_\CO\}$.

A convenient choice of a linear functional $\a$ is a linear combination of
derivatives around the crossing symmetric point
\begin{align}\label{choice}
  \a\Big[f(z)\Big] = \sum_{n=0}^N \a^n  \left.  \frac{\partial^n}{\partial z^n} f(z) \right|_{z=1/2}\ ,
\end{align}
where we can take an arbitrary number for $N$.
With this choice, one can translate the problem (\ref{constraint})
into a semi-definite programming problem\cite{Poland:2011ey}.

To see this, we first note that
the derivatives of the Virasoro conformal blocks can (approximately) factorize into
two pieces
\begin{align} \label{factorize2}
  \frac{\partial^n}{\partial z^n} \CG(c,h_\phi,h_\CO; 1/2) =
  \chi(c,h_\CO)\cdot \CP_n(c,h_\phi,h_\CO)\ ,
\end{align}
one of which is a positive function of conformal weight $h_\CO$ ($h_\CO\in\mathbb{R}^+$), $\chi(h_\CO)$,
and the other is a polynomial of $h_\CO$, $\CP_n(h_\CO)$.
The precise form of the positive function $\chi(h_\CO)$ will be presented later.
It is noteworthy that the function
$\chi(h_\CO)$ is independent of the order of derivative $n$, and thus we can
reduce the problem (\ref{constraint}) into a nicer one of finding a real $N$-dimensional vector $\vec \a$
such that
\begin{align}\label{constraint2}
  \sum_{n=0}^N \a^n \partial_n \CG(c,h_\phi,0;1/2) > 0\ , \qquad
  \sum_{n=0}^N \a^n \CP_n(c,h_\phi,h_\CO) \geq 0 \ ,
\end{align}
for any positive $h_\CO$. Using the Hilbert theorem, one can further reduce (\ref{constraint2})
into a semi-definite programming (SDP)\cite{Poland:2011ey}:
find $\{\vec \a \in\mathbb{R}^N, \CB \succeq0, \CC \succeq 0\}$ such that
\begin{gather}
  \sum_{n=0}^N \a^n \partial_n \CG(c,h_\phi,0;1/2) > 0\ ,
  \nonumber \\
  \sum_{n=0}^N \a^n \CP_n(c,h_\phi,h_\CO)  = [h_\CO]_{\d_1}^T \cdot \CB \cdot [h_\CO]_{\d_1} +
  h_\CO [h_\CO]_{\d_2}^T \cdot \CC \cdot [h_\CO]_{\d_2}  \ , \label{SDPP2}
\end{gather}
where $M \succeq 0$ denotes a positive semi-definite matrix and $[h_\CO]_\d^T=(1,h_\CO,h_\CO^2,...,h_\CO^\d)$.
Here $\d_1=\lfloor d/2 \rfloor$ and $\d_2=\lceil (d-1)/2 \rceil$ where $d$ is the degree of
$\CP_n(c,h_\phi,h_\CO)$.


\section{Numerical Analysis}

\subsection{Settings}

\paragraph{Input parameters}
In order to perform the numerical bootstraps, we need to set various parameters in the SDP problem
relevant to the extremal CFTs. The central charge $c=24k$, and
the conformal weight of the chiral operator $\phi(z)$ is given by $h_\phi=k+1$.
We also introduce a gap in spectrum between the vacuum and any primary operator $\CO$
so that $h_\CO\geq h_\text{gap}=k+1$. As the extremal CFT has the spectrum of integer conformal weights, we introduced $m_0$ low-lying discrete states of integer conformal weights from $h_{gap}$.

Given the above CFT data, we focus on the feasibility of semi-definite programming problem
(\ref{SDPP2}),
\begin{align} \label{sdp1}
  \sum_{n=0}^N \a^n \partial_n \CG(c=24k,h_\phi=k+1,0;1/2) = 1
\end{align}
and
\begin{align} \label{sdp2}
  \sum_{n=0}^N \a^n \CP_n(c=24k,h_\phi=k+1,h_\CO)  = [h_\CO]_{\d_1}^T \cdot \CB \cdot [h_\CO]_{\d_1} +
  h_\CO [h_\CO]_{\d_2}^T \cdot \CC \cdot [h_\CO]_{\d_2}
\end{align}
with
\begin{align} \label{sdp3}
  h_\CO & = h_\text{gap} + m  \qquad ( m=0,1,2,..,m_0 )
  \nonumber \\
  h_\CO  & = h_\text{gap}+ ( m_0 + 1) +x \qquad ( x \geq 0 ) \ .
\end{align}
Here we use the scale symmetry of the SDP problem (\ref{SDPP2}) to fix
the normalization (\ref{sdp1}).
If this semi-definite programming is feasible, the bootstrap
equations (\ref{bootstrap1}) cannot be satisfied by any consistent conformal field theory
with the above given data. On the other hand, if such a feasible
solution is failed to be found,
the bootstrap equations (\ref{bootstrap1}) may or may not be satisfied.
Thus one cannot say anything conclusive  on the existence
of the extremal conformal field theory.

\paragraph{Stability} The numerics of Virasoro bootstrap are mainly based
on the Zamolodchikov recursive formula (\ref{recursive}). If we solve
the recursion relation iteratively, the function $H(c,h_\phi,h_\CO,\mathpzc{q}(z))$
can be expressed as a power series expansion in terms of the
elliptic variable $\mathpzc{q}(z)$. However, to set a solvable numerical problem, we need to
approximate the function $H(c,h_\phi,h_\CO,\mathpzc{q}(z))$ by
a polynomial of finite order in $\mathpzc{q}(z)$ and truncate
the number of iteration to solve the recursion relation.

Due to the fact that $R_{1,1}(c,h_\phi)=0$ in (\ref{recursive}),
one can show that the terms of order up to $2p$ in $H(c,h_\phi,h_\CO,\mathpzc{q})$ become exact
after repeating the iteration by $p$-times.
It is therefore legitimate to approximate the function
$H(c,h_\phi,h_\CO,\mathpzc{q}(z))$ by a polynomial of order $2p$ in $\mathpzc{q}(z)$
for the iteration by $p$-times.

\begin{figure}
\begin{center}
\includegraphics[height=2.6in]{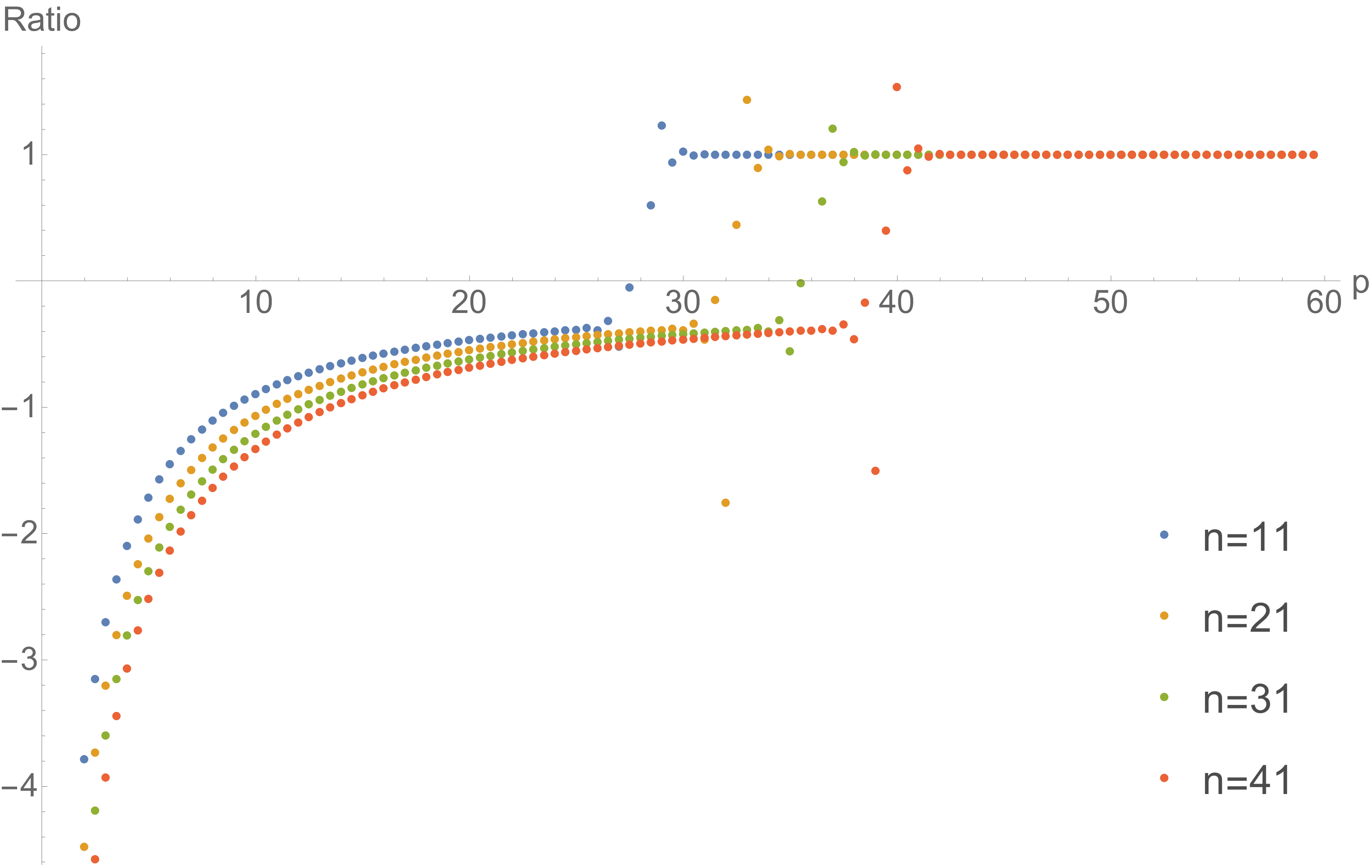}
\end{center}
\caption{This plot shows the stabilization of Taylor coefficients
of the Virasoro conformal block for $k=30$. The vertical axis represents
$\partial_n \CF(c,h_\phi,h_\text{gap},1/2)_{p+1}/\partial_n \CF(c,h_\phi,h_\text{gap},1/2)_{p}$
where the subscripts denote the numbers of iteration performed to obtain the
Virasoro conformal block $\CF(c,h_\phi,h_\text{gap},1/2)$.}\label{stabilityapproxk=10}
\end{figure}

To justify the validity of approximation via the finite number of iteration,
the convergence of series should be guaranteed.
In other words, numerical values of the Virasoro conformal blocks and their derivatives around
the crossing symmetric point $z=1/2$ should approach
certain values as the size of truncation increases.
Figure \ref{stabilityapproxk=10} shows the patterns
of the various Taylor expansion coefficients of a Virasoro conformal
block. The points in this plot represents the ratio
of Taylor coefficients obtained after repeating the iteration
by $(p+1)$-times and $p$-times. For a sufficiently large number of iteration,
this plot shows the ratio finally settles down to $1$, which means
that Taylor coefficients are stabilized.

\paragraph{Details of (\ref{factorize2})} Let us now discuss how we can approximate $\partial_n \CG(c,h_\phi,h_\CO;1/2)$ as $\chi(c,h_\CO) \cdot \CP_n(c,h_\phi,h_\CO)$ where $\chi(c,h_\CO)$ is a positive function of $h_\CO $. As discussed above, we first take a sufficient number of iterations, say $p$, to get an accurate approximation of $H(c,h_\phi,h_\CO,\mathpzc{q}(z))$ as a polynomial of order $2p$ in $q$. According to the recursive formula, each coefficient of the polynomial is proportional to $\prod_{m,n} \frac{1}{h-h_{m,n}(c)}$. It is then straightforward to show that $\chi(c,h_\CO)$ can be expressed as
\begin{align}
\chi(c,h_\CO) =  \frac{\left( 16 \mathpzc{q}(\frac{1}{2}) \right)^{h_\CO}}{\prod_{m,n} h_\CO - h_{m,n}(c)}
\end{align}
where the denominator is the common denominator in $h_\CO$ of $H(c,h_\phi,h_\CO,\mathpzc{q}(z))$.
The fact that $h_{m,n}(c) < 0$ for $c\geq 25$ ensures that
$\chi(h_\CO)$ is positive definite for $h_\CO\geq0$\footnote{We can also show the positivity of $\chi(h_\CO)$ even for $c=24$.}. This allows us to determine the form of the polynomial $\mathcal{P}_n(h_\CO)$ for given $k$.

\subsection{Numerical Results}

To determine if the semi-definite programming (\ref{sdp1}) and (\ref{sdp2}) with (\ref{sdp3}) is feasible or not,
we use the package {\tt SDPB}\cite{Simmons-Duffin:2015qma} with the option ${\tt dualErrorThreshold}=10^{-200}$.
The solver terminates itself if a solution $(\a,\CB,\CC)$ satisfying
its constraints (\ref{sdp1}) and (\ref{sdp2}) to sufficient precision set to $10^{-200}$ is found.
If the {\tt SDPB} finds a feasible solution, there is no CFT with the given CFT data consistent with unitarity and crossing symmetry.
On the other hand, if the {\tt SDPB} fails to find a feasible solution, the existence of a consistent CFT with the input CFT data becomes inconclusive.

We numerically checked the consistency of the conformal bootstrap equation for
various values of central charge $c=24k$ with the derivative order in (\ref{choice})
set to $41$, namely $N=41$.
For $m_0=0$, we do not find a feasible solution up to $k=70$.
We can however constrain the bootstrap equation more strongly by introducing a sufficiently large number of
discrete states of low-lying integer conformal weights, i.e., $m_0\neq0$. Around $m_0 = 2000$,
the {\tt SDPB} start to find feasible solutions $(\a_\ast,\CB_\ast,\CC_\ast)$ for $k=20$--$30$, $40$, $50$, $60$ and $70$
to the semi-defining programming of our interest.
Furthermore, the solutions satisfy the strictly positivity for $h_\CO=k+1$,
\begin{align}\label{resultaaa}
  \sum_{n=0}^N \a_\ast^n \CP_n(c=24k,h_\phi=k+1,h_\CO=k+1) > 0 \ .
\end{align}
\begin{figure}[h]
\begin{center}
\includegraphics[width=.84\textwidth]{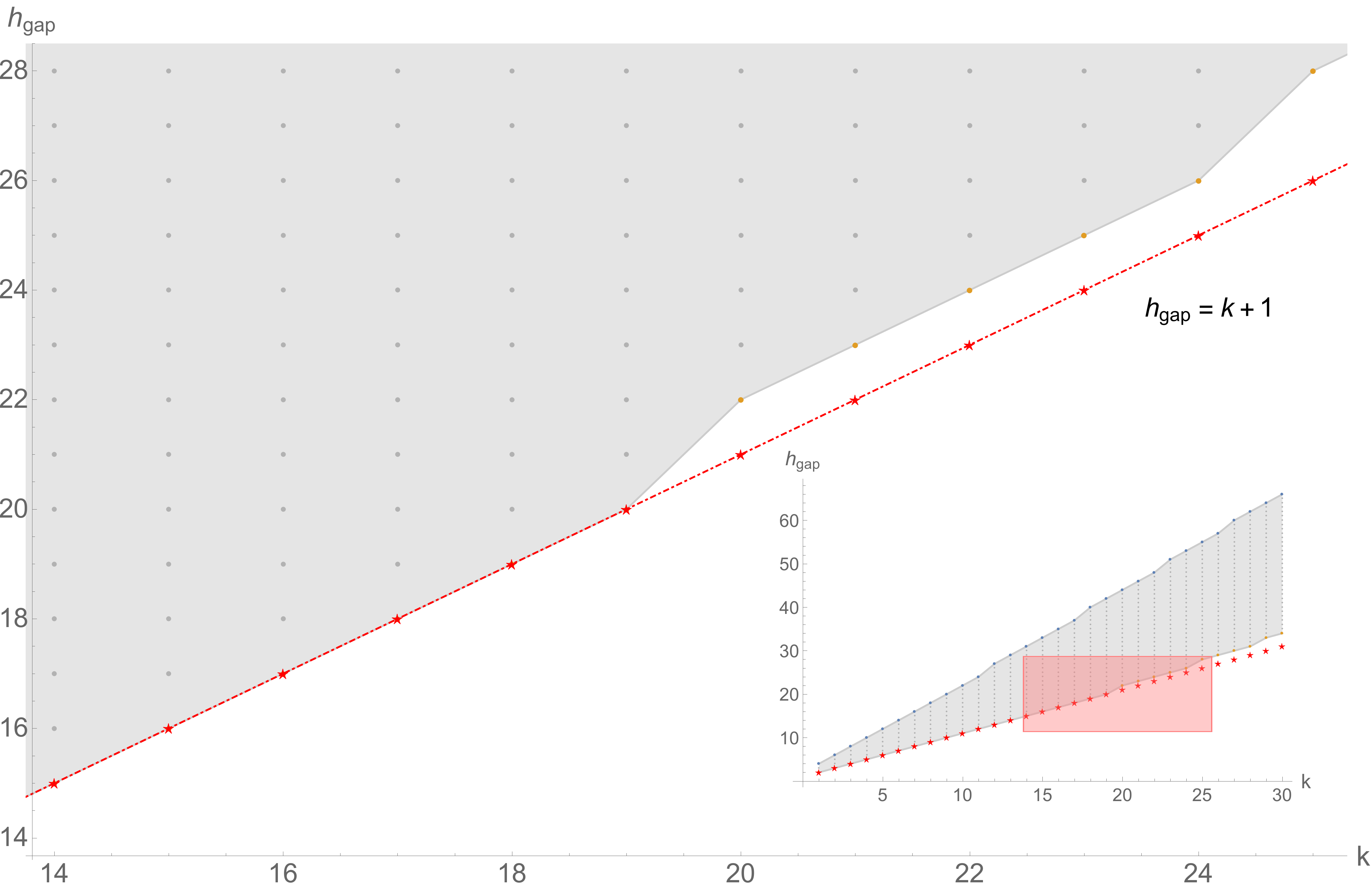}
\end{center}
\caption{The {\tt{SDPB}} solver finds optimal solutions of (\ref{opt1}) with the negative maximum value inside grey region. However, (\ref{opt1}) has no optimal solutions in exterior region.
The red dotted line indicates the case of extremal CFT, i.e., $h_{gap} = k + 1$. In this analysis, $h_{gap}$ takes integer values.
}\label{fig2}
\end{figure}
It turns out that one can check the above result independently by analyzing another numerical problem. The alternative problem is given below,
\begin{align} \label{opt1}
   \text{maximize } &\sum_{n=0}^N \a^n \partial_n \CG(c=24k,h_\phi=k+1,0;1/2)
  \nonumber \\
  \text{such that } &\sum_{n=0}^N \a^n \CP_n(c=24k,h_\phi=k+1,h_\CO=h_\text{gap}) = 1
  \nonumber \\  \hspace*{1.8cm}
  \text{and } &\sum_{n=0}^N \a^n \CP_n(c=24k,h_\phi=k+1,h_\CO)  \geq 0
\end{align}
with
\begin{align}
  h_\CO & = h_\text{gap} + m  \qquad ( m=1,2,..,m_0 )
  \nonumber \\
  h_\CO  & = h_\text{gap}+ ( m_0 + 1) +x \qquad ( x \geq 0 ) \ ,
\end{align}
where $m_0=2000$. This is because a feasible solution to (\ref{sdp1}) and (\ref{sdp2})
where $\sum_{n=0}^{N} \a^n \CP_{n}(c=24k,h_\phi=k+1,h_\CO=h_\text{gap}) > 0$ indeed implies that the above optimal problem
should not have a solution of negative maximum value. We have examined if (\ref{opt1}) has optimal solutions or not by adjusting
the parameter $h_\text{gap}$. In Figure \ref{fig2}, optimal solutions with negative maximum value are found in the shaded region.
On the other hand, no optimal solutions are found outside the shaded region.

The red dotted line, $h_\text{gap} = k+1$, represents the extremal CFTs. This result supports our previous numerical results discussed before (\ref{resultaaa}) suggesting that the non-existence of extremal CFTs for $k=20$--$30, 40, 50, 60$, and $70$ do not exist\footnote{Outside the shaded region, we observed that the {\tt SDPB} finds a (dual) feasible solution to (\ref{opt1}) with positive value for $\a^n \partial_n \CG(c=24k,h_\phi=k+1,0;1/2)$ (a.k.a. {\tt{dual objective}}).}. The extremal CFTs for $k=2$--$19$  however lie in the shaded region, and their existence remains inconclusive .

\section{Conclusions and Discussions}

In this paper, we utilized conformal bootstrap method to analyze the extremal CFT dual to the 3D pure quantum gravity with a negative cosmological constant. This naturally led us to focus on the $c=24 k$ holomorphic CFT which have a large gap
 between the vacuum and the lowest primary state.
We find numerical evidence for the non-existence of the extremal CFTs for $k \ge 20$. This indicates, under the assumption of holomorphic factorization,  the pure gravity in the weakly curved AdS$_3$ does not exist as a consistent quantum theory.

There has been a recent proposal to include ``small black holes'' as additional physical states in the pure quantum gravity in AdS$_3$\cite{Benjamin:2016aww}. These small black holes are supposed to have zero entropy classically but finite entropy due to quantum corrections.
This modification introduces new primary states of conformal weight $h=k$ in  the dual CFT. This new CFT is referred to as near-extremal CFT.
For the near-extremal CFTs, we have repeated the numerical analysis of the semi-definite programming  (\ref{sdp1}), (\ref{sdp2}) and (\ref{opt1}) by setting $h_\phi=h_\text{gap}=k$.
\begin{figure}[t!]
\begin{center}
\includegraphics[width=.84\textwidth]{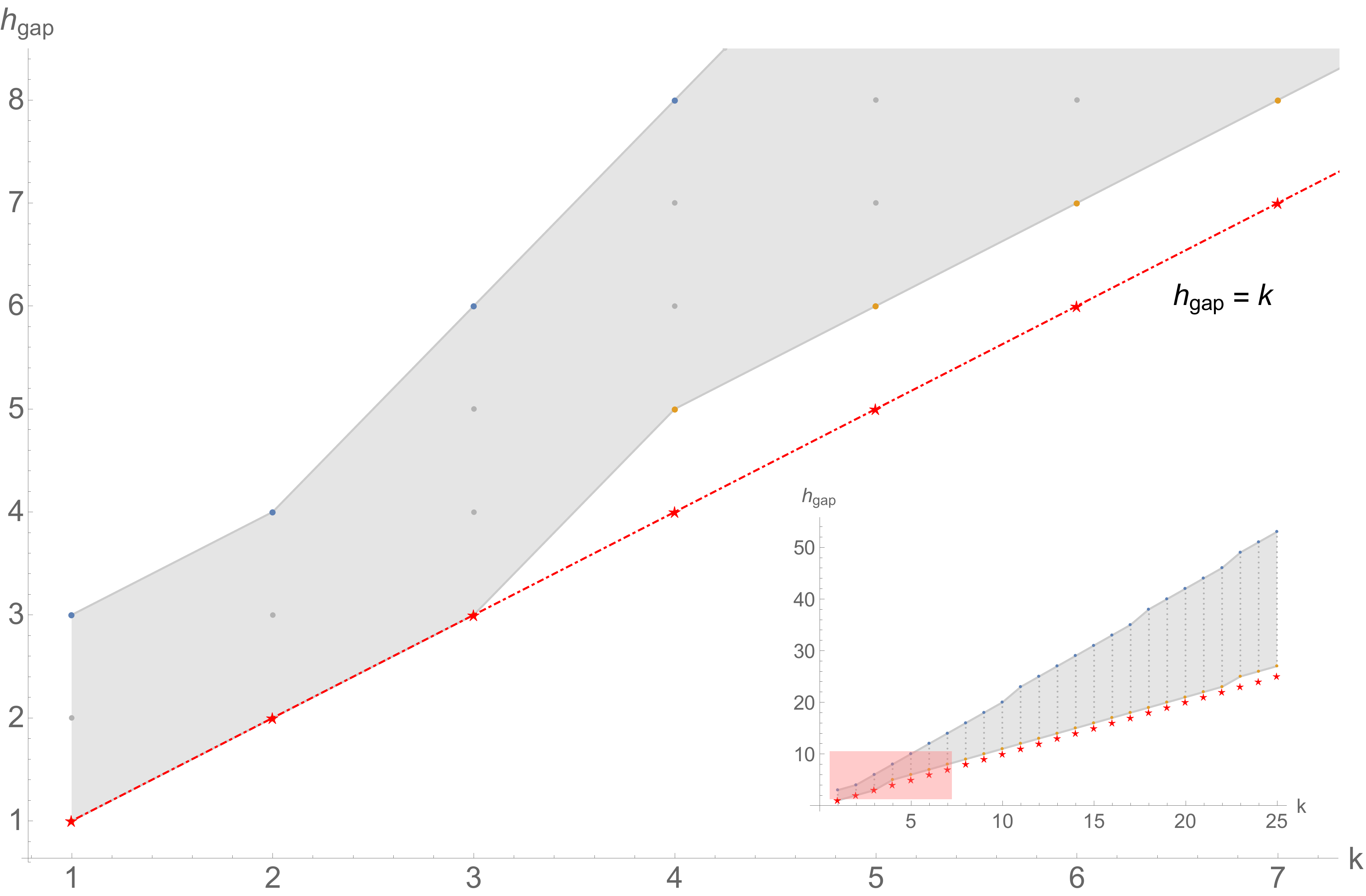}
\end{center}
\caption{The {\tt{SDPB}} solver finds optimal solutions of (\ref{opt1}) with the negative maximum value inside grey region. However, (\ref{opt1}) has no optimal solutions in exterior region.
The red dotted line indicates the case of near-extremal CFT, i.e., $h_{gap} = k $. In this analysis, $h_{gap}$ takes integer values.}\label{resultNE}
\end{figure}

As depicted in Figure \ref{resultNE}, our numerical results indicate non-existence of the CFT for $k \ge 4$. Comparing with the result of the extremal CFTs, a fewer CFTs are allowed to be located in the shaded region.

It is  important to explore  pure quantum gravity
in AdS$_3$ without relying on the holomorphic factorization. One can use the conformal and modular
bootstrap methods to study various constraints on such theories\cite{Hellerman:2009bu,Collier:2016cls}.
The supersymmetric extension of extremal CFTs were explored in \cite{Gaberdiel:2008xb,Harrison:2016hbq}.
We would like to generalize our analysis to investigate
the consistency of such supersymmetric extremal CFTs in large central charge limit.
It would be also interesting to apply the numerical conformal bootstrap to study various features of generic
quantum gravity in higher dimensional AdS space. We leave them for future works.

\section*{Acknowledgments}
We thank J. Harvey, S. Terashima, E. Witten and P. Yi for discussions.
We thank KIAS Center for Advanced Computation for providing computing resources.
S.L. thank the organizers of the workshops ``Nambu Symposium'' at the University of Chicago,
``String-Math Trimester'' at Henri Poincare Institute,
``Boundaries and Defects in Quantum Field Theories'' at Aspen Center for Physics (supported by
National Science Foundation grant PHY-1066293), and
``String and Fields 2016'' at Yukawa Institute for Theoretical Physics for their kind hospitality
during some stages of this work. The work of K.L. is supported in part by the National Research Foundation of Korea (NRF) Grants No. 2006-0093850.

\bibliographystyle{JHEP}
\bibliography{Refs}

\end{document}